\documentclass[twocolumn,superscriptaddress,showpacs,preprintnumbers,amsmath,amssymb,prl]{revtex4}
\usepackage{graphicx}
\usepackage{dcolumn,xcolor,ulem}
\usepackage{amsmath} 
\usepackage{amssymb}
\usepackage{amsfonts}
\usepackage{bm}
\usepackage[latin1]{inputenc}

\newcommand\gp{\dot\gamma}

\newcommand\Rey{\mbox{\textit{Re}}}  
\newcommand\Wei{\mbox{\textit{Wi}}}  
\newcommand\Ta{\mbox{\textit{Ta}}}  
\newcommand\El{\mathcal{E}} 

\begin{document}

\title{Inertio-elastic instability of non shear-banding wormlike micelles}

\author{C.~Perge}
\affiliation{Universit\'e de Lyon, Laboratoire de Physique, \'Ecole Normale Sup\'erieure de Lyon, CNRS UMR 5672, 46 All\'ee d'Italie, 69364 Lyon cedex 07, France}
\author{M.A.~Fardin}
\email{marcantoine.fardin@ens-lyon.fr}
\affiliation{Universit\'e de Lyon, Laboratoire de Physique, \'Ecole Normale Sup\'erieure de Lyon, CNRS UMR 5672, 46 All\'ee d'Italie, 69364 Lyon cedex 07, France}
\affiliation{The Academy of Bradylogists}
\author{S.~Manneville}
\affiliation{Universit\'e de Lyon, Laboratoire de Physique, \'Ecole Normale Sup\'erieure de Lyon, CNRS UMR 5672, 46 All\'ee d'Italie, 69364 Lyon cedex 07, France}
\affiliation{Institut Universitaire de France}

\date{\today}

\begin{abstract}
Homogeneous polymer solutions are well-known to exhibit viscoelastic flow instabilities: purely elastic when inertia is negligible, inertio-elastic otherwise. Recently, shear-banding wormlike micelles solutions were also discovered to follow a similar phenomenology. In the shear-banding regime, inertia is usually negligible so only purely elastic flows have been reported. Here, we investigate a non-shear-banding solution where inertia becomes significant, leading to flow patterns akin to the inertio-elastic regime of dilute polymer solutions. We show that the instability follows a supercritical bifurcation and we investigate the structure of the inertio-elastic vortices that develop above onset.
\end{abstract}

\pacs{83.80.Qr, 47.20.-k, 47.27.-i, 47.50.Gj }
\maketitle


In a simple Newtonian fluid like water flows tend to become unstable and eventually turbulent due to inertia, for increasing values of the Reynolds number $\Rey\equiv\tau_1 \gp$, where $\gp$ is a characteristic velocity gradient or deformation rate. The characteristic time of the fluid is the viscous dissipation time $\tau_1\equiv \rho d^2/\eta$, where $\eta$ is the shear viscosity of the fluid, $\rho$ its density, and $d$ the characteristic length in the velocity gradient direction. Nonetheless, determining the instability threshold and the flow patterns that emerge can be challenging. The difficulty depends on the geometry, in particular whether it is a straight pipe or a curved geometry like in the Taylor-Couette (TC) flow between concentric cylinders~\cite{Schmid:2001}. The latter case is {\it a priori} simpler because the flow is expected to be linearly unstable. More precisely, Taylor~\cite{Taylor:1923} showed that the purely azimuthal base flow becomes unstable when the Taylor number $\Ta= \Lambda^{1/2} \Rey$ exceeds some critical value $\Ta_c$, where $\Lambda$ denotes the dimensionless curvature of the streamlines. When only the inner cylinder of radius $R_i$ is rotating at angular velocity $\Omega$ and for a small gap $d$ between the two cylinders (i.e. $d/R_i \ll 1$), one takes $\gp\simeq \Omega R_i/d$ to compute $\Rey$ and $\Lambda$ is simply given by $\Lambda=d/R_i$. In this case the value of the threshold is $\Ta_c\simeq 41$~\cite{Taylor:1923,Donnelly:1960}. 

While the case of an incompressible isothermal Newtonian fluid has been tackled and understood for about a century, the situation is much less clear when time scales different from the inertial time $\tau_1$ become relevant. For instance, the effects of compressibility, magnetic field, or temperature have been investigated~\cite{Schmid:2001,Chandrasekhar:1961}. Here, we are interested in viscoelastic fluids, where the stress relaxation brings up an additional time scale $\tau_2$. The archetype of a viscoelastic fluid is a polymer solution, where $\tau_2$ can be a Rouse, a Zimm, or a reptation time scale, depending on the degree of overlap between chains~\cite{deGennes:1979}. Within this context another route to turbulence exists, which does not involve inertia and is referred to as ``purely elastic''~\cite{Groisman:2000,Larson:1990,Morozov:2007}. If $\Rey\ll 1$, the Weissenberg number $\Wei \equiv \tau_2\gp$ takes the role of control parameter. In the TC case, Larson \textit{et al.} showed that vortex flows also emerge for $\Ta >\Ta_c$, but with $\Ta = \Lambda^{1/2} \Wei$, and for instance $\Ta_c\simeq 6$ if the fluid follows the Upper Convected Maxwell model (the simplest tensorial viscoelastic model)~\cite{Larson:1990}. If both $\Rey$ and $\Wei$ are large flow instabilities are referred to as ``inertio-elastic.'' In this case, a precise instability scaling is lacking even in the simple TC geometry~\cite{Muller:2008}. Dimensional analysis suggests that the Taylor number should be written as $\Ta=\Lambda^{1/2} f(\Rey , \Wei)$, with  $\lim_{\El\to 0} f(\Rey , \Wei)=\Rey$ and $\lim_{\El\to \infty} f(\Rey , \Wei)=\Wei$, where $\El\equiv \Wei/\Rey =\tau_2/\tau_1$ is the elasticity number. Unmasking the complete scaling of instabilities for any elasticity $\El$ is a challenging task that currently involves several research groups~\cite{Fardin:2014pp}.

Until recently, polymer solutions were the only systems used experimentally in connection to viscoelastic instabilities. Since a few years, surfactant wormlike micelles have also been shown to exhibit elastic instabilities~\cite{Fardin:2009,Fardin:2012d}. Wormlike micellar solutions are well-known model systems for rheological research~\cite{Lerouge:2010}. In the linear rheology regime (i.e. for small deformations) they display essentially Maxwellian behaviour, characterized by an elastic modulus $G_0$ and a single relaxation time $\tau_2$ (see Supp. Fig.~1), in contrast to polymer solutions whose wide spectrum of relaxation times may modify flow instabilities~\cite{Larson:1994}. Moreover wormlike micelles constantly break and recombine~\cite{Cates:2006}, contrary to polymers that can be irreversibly degraded both chemically and mechanically. But maybe even more importantly, semidilute and concentrated wormlike micelles are known to present shear banding in the nonlinear rheology regime: above a critical shear rate, the fluid becomes dramatically shear-thinning, the stress plateaus, and the fluid splits in at least two bands of different local viscosities and shear rates~\cite{Lerouge:2010}. In this case, inertia is negligible and purely elastic instabilities generate vortex flows that are confined in the high shear band~\cite{Fardin:2012d}. 

In recent years, wormlike micelles have been used to probe viscoelastic instabilities in various geometries. In particular, in the cross-slot geometry, a large range of elasticities $\El$ could be spanned by using different surfactant and salt concentrations~\cite{Dubash:2012,Haward:2012} but at the expense of a loose control of the rheology. Some solutions were shear-thickening due to shear-induced structures, while others were shear-banding and some may have been simply shear-thinning; three typical cases encountered in surfactant solutions from dilute to concentrated~\cite{Lerouge:2010}. The interpretation of these data is then complicated by the possible interplay between elasticity and inertia, as well as other rheological factors. In the TC geometry, only shear-banding~\cite{Fardin:2012d} and shear-thickening~\cite{Fardin:2013pp} systems have been probed so far. Here, we show for the first time that a well-characterized {\it non} shear-banding (homogeneous) wormlike micellar solution with $\El\sim 1$ displays a supercritical inertio-elastic instability similar to polymer solutions. By means of ultrasonic imaging, we also unveil a previously unreported asymmetric structure in the inertio-elastic vortices that develop above onset.

\begin{figure}
\centering
\includegraphics[width=7cm,clip]{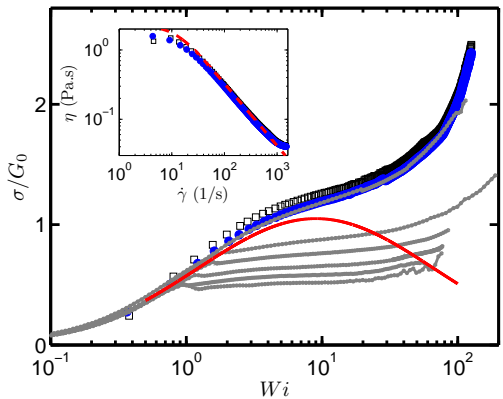}
\caption{Dimensionless flow curves $\sigma/G_0$ vs $\Wei=\gp\tau_2$ of our wormlike micellar system with [CTAB]=0.1~M seeded with 1\%~w/w hollow glass spheres ($\square$), seeded with 1\%~w/w Kalliroscope platelets (blue $\bullet$), and free of any tracers (gray dots) compared with shear-banding solutions with [CTAB]=0.2, 0.3, 0.4, 0.5 and 0.6~M (gray dots from top to bottom). All flow curves correspond to samples with [NaNO$_3$]=0.3~M at $T=30^\circ$C. The red line shows the approximate frontier between shear-banding flows (below the line) and non shear-banding flows (above). The flow curves in gray dots are reproduced from Ref.~\cite{Fardin:2012c}. Inset: viscosity of the [CTAB]=0.1~M sample vs $\gp$. The red dashed line is a fit to the Carreau model (see text).
\label{fig1}}
\end{figure} 

Our wormlike micellar system is made of 0.1~M cetyltrimethylammonium bromide (CTAB) and 0.3~M sodium nitrate (NaNO$_3$) in water. The temperature is held fixed at $T=30\pm 0.1^\circ$C. This system was specifically chosen because, as shown in Fig.~\ref{fig1}, it stands out as the ``first non-shear-banding'' solution in a family of systems at the same temperature and salt concentration but with increasing surfactant concentrations~\cite{Fardin:2012c}. At this temperature and salt concentrations, solutions with [CTAB]=0.2--0.7~M are known to be shear-banding and to exhibit a purely elastic instability of the high shear band~\cite{Fardin:2012c}. In contrast, the present 0.1~M sample is only shear-thinning, with a viscosity well described by the Carreau model, $\eta(\gp)=\eta_0(1+(\tau_2 \gp)^2)^{-n}$. We found $\tau_2=0.08$~s and $G_0=28$~Pa based on small angle oscillatory shear (see Supp. Fig.~1), i.e. $\eta_0=G_0\tau_2=2.2$~Pa.s, and a shear-thinning index of $n=0.45$ (see inset of Fig. 1). We can then compute $\Rey=\rho d^2\gp /\eta(\gp)$ using $\rho=10^3$ kg.m$^{-3}$, and $\Wei =\tau_2 \gp$. As commonly done for shear-thinning polymer solutions~\cite{Dutcher:2013}, we assume that $\tau_2$ is independent of $\gp$. We checked that taking $\Wei$ to be the ratio $N_1/\sigma$ of the first normal stress difference to the shear stress yields similar estimates for $\Wei$ (see Supp. Fig.~2). 

\begin{figure*}
\centering
\includegraphics[width=11.5cm,clip]{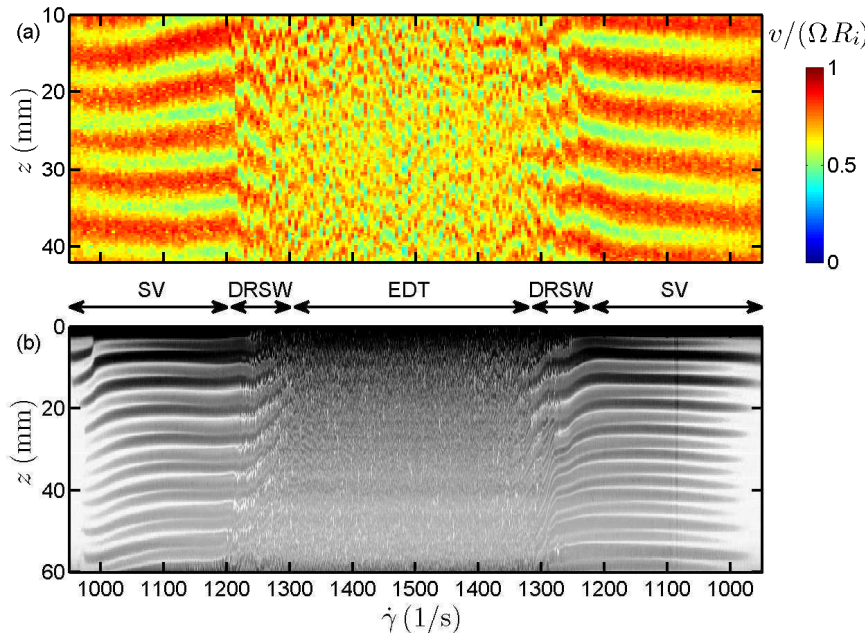}
\caption{Spatiotemporal diagrams of the flow states reached for a linear ramp in shear rate first increasing from $\gp=950$ to $1500$~s$^{-1}$ and then decreasing from $\gp=1500$ down to $950$~s$^{-1}$. (a)~Velocity $v(r_0,z)/(\Omega R_i)$ coded in linear color levels as measured using ultrasonic imaging  at a distance $r_0=d/4$ from the rotating inner cylinder in the sample seeded with hollow glass spheres; the ramp rate is 2.5~s$^{-1}/$s. (b)~Direct visualization of the sample seeded with Kalliroscope platelets; the ramp rate is 1.7~s$^{-1}/$s. See supplementary material for more details on the construction of these diagrams. The flow takes the following route to turbulence: CF $\rightarrow$ SV $\rightarrow$ DRSW $\rightarrow$ EDT and back again (see text).
\label{fig2}}
\end{figure*} 

In all experiments, the fluid is sheared in a TC device with inner rotating cylinder adapted to a rheometer (TA Instruments ARG2). The height of the device is $H=60$ mm and the gap width $d=2$ mm, such that we can assume no strong impact of the top and bottom boundaries. The top of the device is sealed by a plug to prevent evaporation of the sample and its possible ejection at high rotation rates. The inner radius is $R_i=23$ mm, so that we can use the small gap approximation ($\Lambda \simeq 0.087 \ll 1$). Flow visualization is performed by seeding the sample with 1\%~w/w microscopic platelets (Kalliroscope AQ-RF), lighting the transparent Plexiglas cell with a LED backlight source (PHLOX LEDW-BL-200$\times$200), and recording movies of the sheared fluid with a standard webcam. Furthermore, two-dimensional ultrasonic velocimetry~\cite{Gallot:2013} is performed in the same device on samples seeded with 1\%~w/w hollow glass beads (Potters Sphericel, mean diameter $6~\mu$m, mean density 1.1), which yields time-resolved maps of the velocity $v(r,z)=v_\theta(r,z)+\frac{v_r(r,z)}{\tan\phi}$, where $r$ is the radial distance to the rotor, $z$ the vertical position, $v_\theta$ and $v_r$ the azimuthal and radial velocity components respectively, and $\phi\simeq 10^\circ$ the angle between the acoustic axis and the normal to the outer cylinder (since $v_r\ll v_\theta$, the measured velocity is roughly the azimuthal velocity--see Fig.~\ref{fig3} and supplemental material for more details). As checked in Fig.~\ref{fig1} and in Supp. Fig.~1, tracer particles used either for flow visualization or for ultrasonic imaging do not significantly alter the rheological properties of our wormlike micellar solution both in the linear and nonlinear regimes.

\begin{figure}
\centering
\includegraphics[width=6.5cm,clip,trim = 0mm 0mm 0mm -5mm]{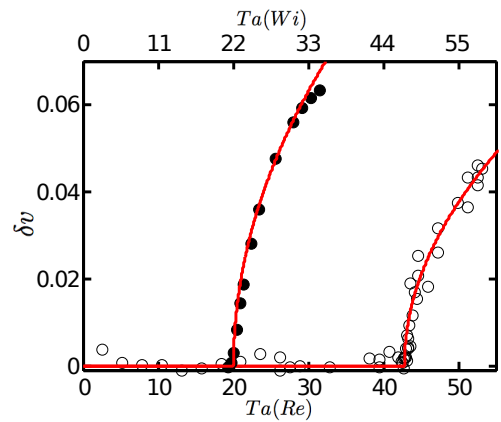}
\caption{Comparison of the bifurcation between CF to TVF in water ($\circ$) and CF to SV in our sample with [CTAB]=0.1~M ($\bullet$). The order parameter is the dimensionless deviation from the base flow $\delta v=\frac{\langle\vert v-v_{\rm lam}\vert\rangle}{\Omega R_i}$, where $v_{\rm lam}$ is the purely azimuthal Couette flow, $v$ the measured velocity field in the gap, and the average is taken on both the radial and vertical directions. The purely inertial Taylor number axis (bottom) is $\Ta (\Rey)=\Lambda^{1/2}\Rey$ and applies to all data points, whereas the purely elastic Taylor number axis (top) is $\Ta (\Wei)=\Lambda^{1/2}\Wei$ and only apply to the inertio-elastic data points ($\bullet$). Red solid lines are power-laws $\sim (\Ta-\Ta_c)^{1/2}$.   
\label{fig4}}
\end{figure}

\begin{figure}
\centering
\includegraphics[width=7cm,clip]{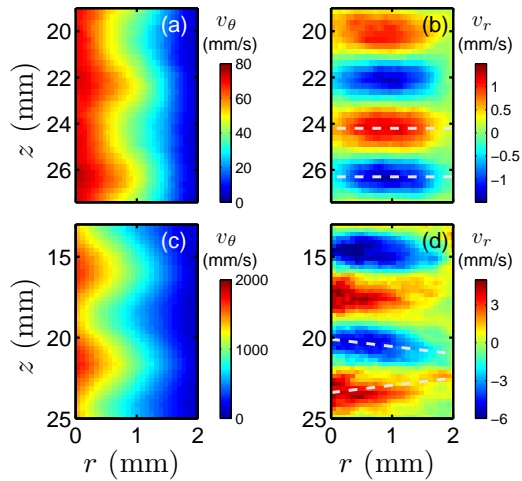}
\caption{Comparison of (a,b) inertial Taylor vortices (TVF) in water at $\gp=40$~s$^{-1}$ and (c,d) the inertio-elastic standing vortices (SV) in our sample with [CTAB]=0.1~M at $\gp=1000$~s$^{-1}$. (a,c) and (b,d) respectively show the azimuthal ($v_\theta$) and radial ($v_r$) velocity maps deduced from ultrasonic imaging as explained in the text. White dashed lines are guides for the eye.
\label{fig3}}
\end{figure}

Figure~\ref{fig2} shows the flow patterns observed as seen both through ultrasonic imaging [Fig.~\ref{fig2}(a)] and direct visualization [Fig.~\ref{fig2}(b)] for an increasing and successively decreasing slow ramp (see also Supp. Movie). No strong hysteresis is observed: the flow goes from the purely azimuthal Couette flow (CF) below $\gp\simeq 950$~s$^{-1}$, to a steady then unsteady vortex flow, and eventually to a turbulent state and back again. The first transition is not seen in Fig.~\ref{fig2} but is studied in detail in Fig.~\ref{fig4}. This sequence of instabilities is strikingly  similar to the case of dilute polymer solutions with moderate elasticity~\cite{Dutcher:2013}. Thus, using the most recent nomenclature, we can call the steady vortex flow a standing vortex (SV) state, that is a Taylor-vortex flow (TVF)~\cite{Andereck:1986} modified by elasticity (such that the boundary between inflow and outflow on Kalliroscope images is sharp~\cite{Dutcher:2013}). The unsteady vortex case observed for $\gp\simeq 1200$ to 1300~s$^{-1}$ is similar to the disordered rotating standing waves pattern (DRSW), the equivalent of wavy vortex flow (WVF)~\cite{Andereck:1986} for moderate elasticity~\cite{Dutcher:2013}. Finally, the last state reached for $\gp\gtrsim 1300$~s$^{-1}$ can be called elastically dominated turbulence (EDT)~\cite{Dutcher:2013}, similar to the disordered oscillations (DO) of Groisman and Steinberg~\cite{Groisman:1996} or to the ``elastic turbulence'' of Giesekus~\cite{Muller:2008}. Calling this state inertio-elastic turbulence (IET) may be more appropriate since $\El\sim 1$, as checked in Fig.~\ref{fig4}. A proper statistical analysis of this turbulent state is left for a future study.

Velocity mapping allows for a precise determination of the first instability bifurcation leading from CF to SV. Figure~\ref{fig4} shows that this first inertio-elastic instability is a linear instability (explaining the absence of hysteresis), given by a pitchfork forward bifurcation like the inertial instability between CF and TVF (illustrated here in pure water seeded with 1\%~w/w hollow glass spheres). The instability diagram is computed by taking the amplitude (or ``order parameter'') as the velocity difference between the measured flow field $v$ and the base azimuthal flow field $v_{\rm lam}\simeq\Omega R_i-\gp r$, normalized by $\Omega R_i$. In the purely inertial case, the control parameter is $\Ta=\Lambda^{1/2}\Rey$ and we indeed recover the classical result $\Ta_c\simeq 41$~\cite{Taylor:1923,Donnelly:1960}. In our wormlike micellar system, 
the thresholds obtained by using either $\Rey$ or $\Wei$ to define $\Ta$ are respectively $\Ta_c(\Rey)\simeq 20$ and $\Ta_c(\Wei)\simeq 22$ so that $\El=\Ta_c(\Wei)/\Ta_c(\Rey)\simeq 1.1$ at instability onset. This value clearly points to the inertio-elastic nature of the instability. Of course, we expect the inertio-elastic threshold to actually depend on both $\Rey$ and $\Wei$: if $\Ta=\Lambda^{1/2} f(\Rey , \Wei) > \Ta_c$, then the flow is unstable. Future theoretical studies should provide an analytical form for $f$ in order to reproduce our results.


In contrast to studies based only on flow visualization, our ultrasonic imaging technique also allows for a quantitative comparison of the SV state (for $\El\sim 1$) and the TVF state (for $\El= 0$). Indeed, as shown in Ref.~\cite{Perge:2013} on a shear-banding flow, one may recover the two horizontal flow components $v_\theta$ and $v_r$ by combining the velocities $v_+$ and $v_-$ from two ultrasonic measurements at the same shear rate but with opposite rotation directions $+\gp$ and $-\gp$ and by using $v_\theta=(v_+-v_-)/2$ and $v_r=\tan\phi\,(v_++v_-)/2$.
Figure~\ref{fig3}(a) and (c) compare the azimuthal velocity maps for water at $\gp=40$~s$^{-1}$ (in the TVF state) and for our sample at $\gp=1000$~s$^{-1}$ (in the SV state). In both cases, vertical oscillations are observed, which are the signature of counter-rotating streamwise vortices that bring slow moving fluid inward in regions of centripetal radial flow and fast moving fluid outward in regions of centrifugal radial flow~\cite{Gallot:2013}. However, the inertial TVF state and the inertio-elastic SV state have slightly different wavelengths, $\lambda/d\simeq2$ and $3$ respectively [note the different vertical scales in Fig.~\ref{fig3}(a,b) and in Fig.~\ref{fig3}(c,d)], and iso-velocity lines are more compressed in the inertio-elastic case. Figure~\ref{fig3}(b) and (d) show the (much smaller) radial velocity maps confirming that for each wavelength of the undulations in the azimuthal velocity, there is a pair of counter-rotating vortices. In contrast to the purely inertial case, the vortex pairs in the SV inertio-elastic case are skewed, with regions of inward velocity pointing downward and regions of outward velocity pointing upward [see white dashes in Fig.~\ref{fig3}(d)]. Note also that the inward (negative) velocity intensity is larger than the outward (positive) one in the SV state [see color map of Fig.~\ref{fig3}(d)]. Finally, instead of being co-localized with the maxima of the oscillations of $v_\theta$ (like for $\El=0$), inward and outward flows in the SV state are shifted by $\lambda/2$ with respect to the inertial case. 

To conclude, we have evidenced for the first time that non shear-banding wormlike micelles are subject to an inertio-elastic instability akin to that known in polymer solutions. This instability appears as supercritical and the resulting Taylor-like vortex flow shows a striking asymmetric structure that is characteristic of elastic vortices. Future experiments will focus on secondary instabilities far above onset, on the route to inertio-elastic turbulence, and on the statistics of such a turbulence. The present results clearly call for a general theoretical framework that would include both inertia and elasticity in minimal models of complex fluids in order to predict at least the first threshold of inertio-elastic instabilities.

M.-A.F. and C.P. contributed equally to this work. The authors would like to thank L.~Casanellas, T.~Divoux, S.~Haward, S.~Lerouge, S.~Muller and G.~McKinley for helpful discussions. This work was funded by the European Research Council under the European Union's Seventh Framework Programme (FP7/2007-2013) / ERC grant agreement n$^\circ$~258803. 

\footnotesize{

\providecommand*{\mcitethebibliography}{\thebibliography}
\csname @ifundefined\endcsname{endmcitethebibliography}
{\let\endmcitethebibliography\endthebibliography}{}

\clearpage
\newpage
\setcounter{figure}{0}
\setcounter{section}{0}

\section{Supplemental material}
\begin{center}
{\bf Inertio-elastic instability of\\non shear-banding wormlike micelles}
\end{center}

\section{Technical details on ultrasonic imaging}

Our TC geometry is equipped with a recently developed two-dimensional ultrasonic velocimetry technique~\cite{Gallot:2013} that allows for the simultaneous measurement of 128 velocity profiles over 30~mm along the vertical direction in the TC geometry~\cite{Gallot:2013}. Ultrafast plane wave imaging~\cite{Sandrin:2001} is used to collect ultrasonic images of the distribution of small acoustic contrast agents seeding the fluid, namely 1\%~w/w hollow glass beads (Potters Sphericel, mean diameter $6~\mu$m, mean density 1.1). Cross-correlation of successive images lead to time-resolved maps of the component $v_y(r,z)$ of the velocity vector, $\mathbf{v}=(v_r,v_\theta,v_z)$ in cylindrical coordinates, projected along the acoustic propagation axis $y$ as a function of the radial distance $r$ to the rotor and of the vertical position $z$. Depending on the shear rate, the time interval between two images can be as low as 50~$\mu$s (see Ref.~\cite{Gallot:2013} for full technical details).

The acoustic axis $y$ is horizontal and makes an angle $\phi\simeq 10^\circ$ with the normal to the outer cylinder so that $v_y=\cos\phi\, v_r + \sin\phi\, v_\theta$. We define the measured velocity as $v=\frac{v_y}{\sin\phi}=v_\theta+\frac{v_r}{\tan\phi}$, which coincides with the azimuthal velocity $v_\theta$ only in the case of a purely azimuthal flow $\mathbf{v}=(0,v_\theta,0)$. More generally $v$ combines contributions from both azimuthal and radial velocity components. Nevertheless, close to instability onset, secondary flows are usually much weaker than the main flow, such that $v\simeq v_\theta$ as can be checked in Fig. 4 of the paper. 
 
\section{Supplemental Movie}
Sup. Movie corresponds to Fig. 2 of the paper. The top part corresponds to the USV data. On the left is the map of the velocity $v(r,z;\gp(t))$ measured in the gap of the TC cell at the location of the ultrasonic beam. The dotted line at $r_0=d/4$ gives the location chosen to draw the spatiotemporal diagram displayed on the right: $v(\gp(t),z;r_{0})$. On the diagram, the dotted line gives the time corresponding to the velocity map. The bottom part corresponds to the data obtained with Kalliroscope platelets. On the left is the section of the outer cylinder illuminated by the LED backlight source, and on the right is the spatiotemporal diagram. The dashed lines fill the same purpose as in the top part.     

\pagebreak
~

\section{Supplemental Figures}

\begin{figure}[h!]
\centering
\includegraphics[width=7cm,clip]{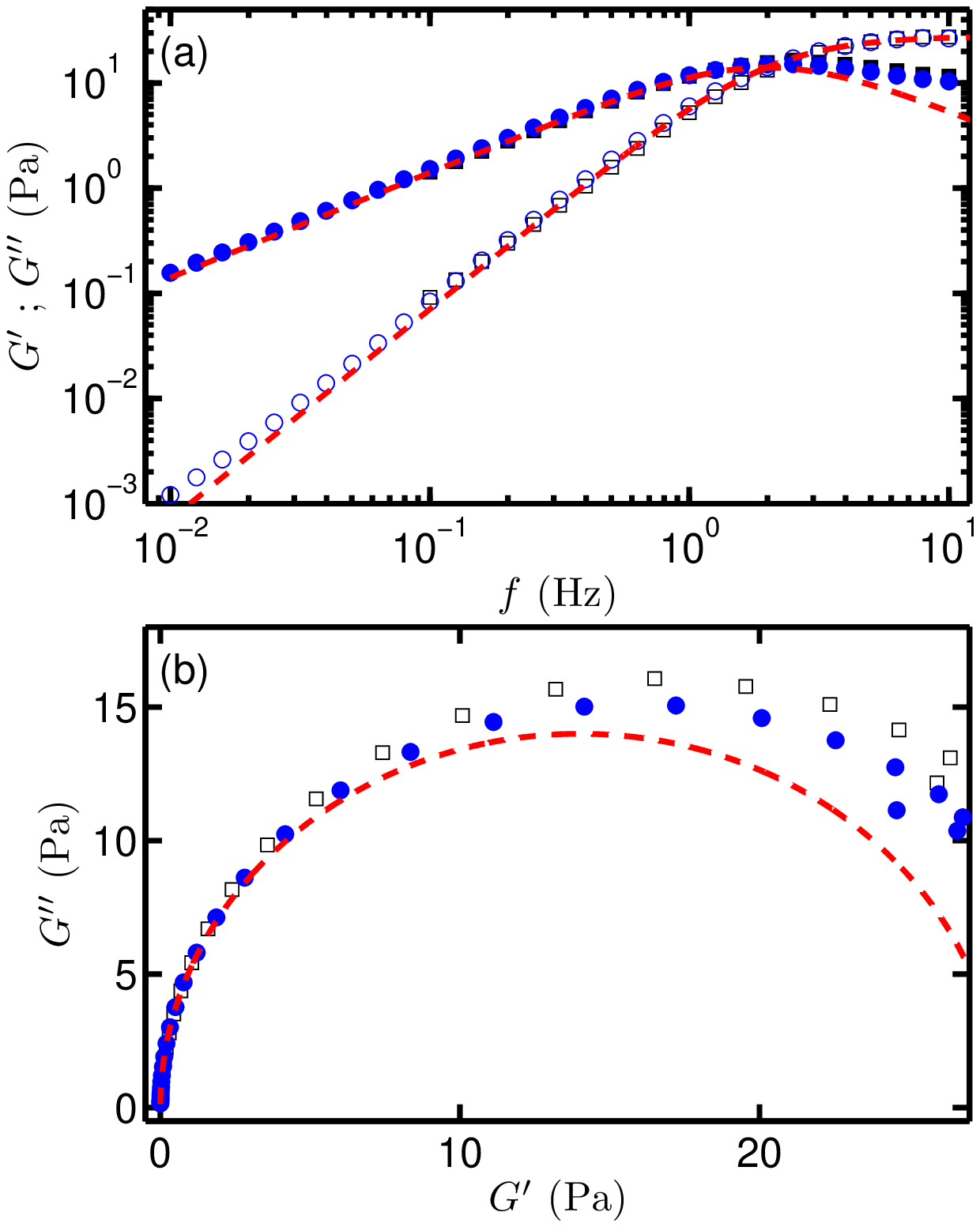}
\caption{(a) Vicoelastic moduli $G'$ (filled symbols) and $G''$ (open symbols) as a function of frequency $f$ for a strain amplitude of 0.1\% and (b) Cole-Cole representation $G''$ vs $G'$ for our wormlike micellar system with [CTAB]=0.1~M seeded with 1\%~w/w hollow glass spheres ($\square$) and seeded with 1\%~w/w Kalliroscope platelets (blue $\bullet$). The red dashed line shows the Maxwell model with a relaxation time $\tau_2=0.08$~s and an elastic modulus $G_0=28$~Pa.
\label{supfig1}}
\end{figure} 

\begin{figure}[h!]
\centering
\includegraphics[width=7cm,clip]{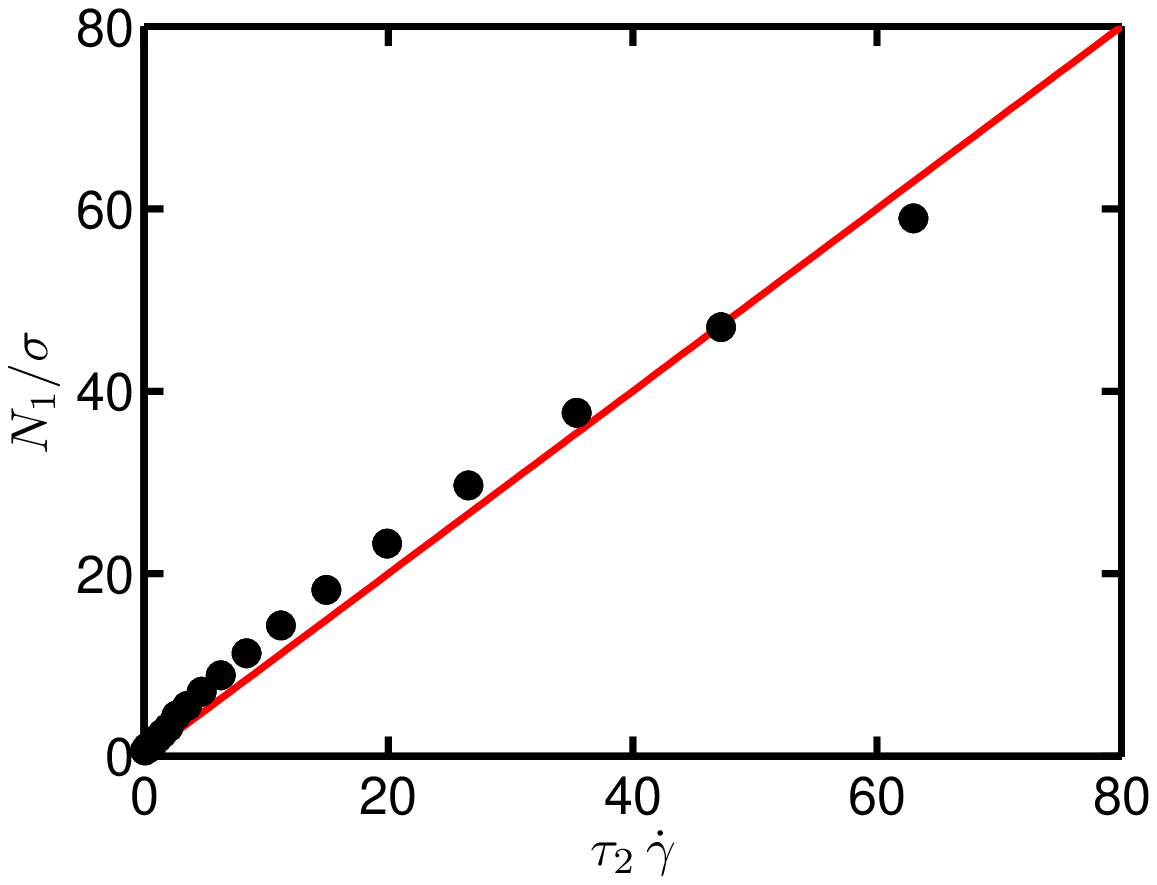}
\caption{Weissenberg number determined as the ratio of the first normal stress difference $N_1$ to the shear stress $\sigma$ as a function of $\tau_2\gp$ in our wormlike micellar system with [CTAB]=0.1~M seeded with 1\%~w/w hollow glass spheres. The red line is $N_1/\sigma=\tau_2\gp$. Experiment performed in a cone-and-plate geometry of diameter 40~mm and cone angle 2~$^\circ$.
\label{supfig2}}
\end{figure} 

\end{document}